\documentclass[useAMS,usenatbib]{mn2e}

\usepackage{mathptmx}
\usepackage{graphicx}
\usepackage{amssymb}
\title[Sloan 2]{The history of star formation and mass assembly in early-type galaxies}

\author[Clemens et al.]{M.S. Clemens$^{1}$, A. Bressan$^{1,2,3}$, B. Nikolic$^{4}$, R. Rampazzo$^{1}$\\
$^{1}$INAF-Osservatorio Astronomico di Padova, Vicolo dell'Osservatorio, 5, 35122 Padova, Italy.\\
$^{2}$SISSA-ISAS, International School for Advanced Studies, via Beirut 4, 34014 Trieste, Italy\\
$^{3}$INAOE, Luis Enrique Erro 1, 72840, Tonantzintla, Puebla, Mexico\\
$^{4}$Astrophysics Group, Cavendish Laboratory, Madingley Road, Cambridge CB3 0HE, UK}
\begin{document}

\date{Accepted. Received; in original form }

\pagerange{\pageref{firstpage}--\pageref{lastpage}} \pubyear{}

\maketitle

\label{firstpage}

\begin{abstract}
We define a volume limited sample of over 14,000 early-type galaxies (ETGs) selected from data release six of the Sloan Digital Sky Survey. The density of environment of each galaxy is robustly measured. By comparing narrow band spectral line indices with recent models of simple stellar populations (SSPs) we investigate trends in the star formation history as a function of galaxy mass (velocity dispersion), density of environment and galactic radius. We find that age, metallicity and $\alpha$-enhancement all increase with galaxy mass and that field ETGs are younger than their cluster counterparts by $\sim 2\;\rm Gyr$. We find negative radial metallicity gradients for all masses and environments, and positive radial age gradients for ETGs with velocity dispersion over $180\;\rm km\,s^{-1}$. Our results are qualitatively consistent with a relatively simple picture for ETG evolution in which the low-mass halos accreted by a proto-ETG contained not only gas but also a stellar population. This fossil population is preferentially found at large radii in massive ETGs because the stellar accretions were dissipationless. We estimate that the typical, massive ETG should have been assembled at $z \lesssim 3.5$. The process is similar in the cluster and the field but occurred earlier in dense environments. 
\end{abstract}

\begin{keywords}
galaxies: elliptical and lenticular, cD, galaxies: formation, galaxies: abundances
\end{keywords}

\section{Introduction}

Observational determinations of the history of star formation in Early-Type Galaxies (hereafter ETGs) are of great importance because hierarchical models of galaxy formation make firm predictions for the relation between age, metallicity and $\alpha$-enhancement as a function of mass. These scaling relations, plotted against the observational proxy for mass, the velocity dispersion (hereafter $\sigma$), have been the focus of many recent studies of ETGs (Annibali et al. 2007, Bernardi et al., 2006, de la Rosa et al. 2007, Gallazzi et al. 2006, Jimenez et al. 2007, Kuntschner et al. 2001/2, Lucey et al. 2007, Mateus et al. 2007, Nelan et al. 2005, Proctor et al. 2004/8, S{\'a}nchez-Bl{\'a}zquez et al. 2006, Smith et al. 2007, Terlevich \& Forbes 2002, Thomas et al. 2005).

These studies clearly show that the simple picture of early formation of low mass galaxies, which then merge to form more massive systems is incorrect. The oldest stellar populations are found in the most massive galaxies -- one aspect of so-called ``downsizing''. However, the observational constraint is sensitive to the epoch at which star formation ceased, not when it started, so that low mass ETGs can still be ``old'', as dynamically bound objects, but have a \emph{mean} stellar age that is much younger. However, this is not sufficient to save the simple hierarchical picture because the $\alpha$-enhancement is also seen to increase with mass - implying more rapid formation for massive objects. This is a prediction of monolithic collapse models. It is important to remember that both the star formation history and mass assembly history of ETGs determine their evolution. 

The most common method of determining the age, metallicity and $\alpha$-enhancement of ETGs is by comparison of the narrow band absorption line indices with simple stellar population (SSP) models. 
It is well-known that stellar population parameters based on these indices are also sensitive to minor episodes of recent star formation. Luminosity-weighted, SSP equivalent stellar population parameters, such as those discussed here, do not therefore, distinguish between a genuinely young galaxy and an old galaxy that has experienced a ``rejuvenation'' event.

Some recent studies, based on the colour-magnitude relation of ETGs using colours that are extremely sensitive to recent star formation, in fact paint a surprising picture of ETG evolution. Schawinski et al. (2007) have used GALEX ultraviolet imaging to show that 30\% of massive ETGs show \emph{ongoing} star formation and that this fraction is higher in low-density environments. A similar picture is given by mid-infrared Spitzer data. Both Clemens et al. (2008) and Bressan et al. (2006) find that $\sim 30\%$ of ETGs in the Coma and Virgo clusters have experienced some star formation in the recent past.

A recent study by Rogers et al. (2007) has combined SDSS spectra and GALEX data to conclude that ``weak episodes of recent star formation'' are a phenomenon more commonly associated with ETGs in the \emph{cluster} environment, a result seemingly, but not necessarily, inconsistent with several studies that find older SSP-equivalent ages in denser environments.

Here we repeat the analysis carried out in Clemens et al. (2006, hereafter Paper~I), which used 3614 objects selected from data release 3 (DR3) of the SDSS. Applying the same selection criteria to DR6 we define a sample of 14353 ETGs, four times as many objects.

\section{Sample selection and data analysis}

Sample selection is identical to that described in Paper~I. The local environmental density is defined as the inverse of the distance to the fifth nearest neighbour, $1/r_5$, corrected for the redshift dependent effect of survey boundaries.

We measure the 21 line-strength indices of the original Lick-IDS system plus the additional indices H$\gamma$F, H$\delta$F, B4000 and HK. However, before measuring the narrow-band indices from each of the SDSS, spectra we smooth the spectra to the wavelength dependent resolution of the Lick-IDS spectra. This step is essential as the models that we use to derive the age, metallicity and $\alpha$--enhancement are based on the Lick system. The index values are then corrected for the smoothing effects of the galaxy's velocity dispersion and aperture corrected to a standard normalized radius (a fraction of the half-light radius). See Paper~I for a more detailed description.

Besides being based on DR6, the present work differs from that of Paper~I only in 2 respects. 
Firstly, in Paper~I we chose to correct for the fixed angular diameter of
$3^{\prime\prime}$ sampled by the SDSS fibre using the radial index measurements of 50
nearby E and S0 galaxies (Rampazzo et al. 2005). Here, we make use of the r-band effective radii
provided in the SDSS catalogue to derive the aperture correction directly.
Statistically, a more reliable correction should be obtained in this way. However, this
also means that we must choose a larger standard radius to which to correct ($r_e$/10 was
used in Paper~I). Typically, the $3^{\prime\prime}$ diameter SDSS fibre samples
$\sim r_e/2$, with very few objects being so large that the fibre samples $r_e/10$. To
avoid extrapolating beyond measured radii we therefore aperture correct to a standard
radius of either $r_e/2$ or $r_e/4$.

A plot of index value versus the fraction of $r_e$ sampled by the fibre will show a
gradient. However, the gradient is due not only to the radial gradients within each
galaxy but also to the correlation between $r_e$ and $\sigma$. Because index values
are correlated with $\sigma$ this effect increases the magnitude of the gradient. To
determine the aperture effect we therefore consider the variation of index value as a
function of the fraction of $r_e$ sampled by the fibre in restricted bins of
$\sigma$. In this way we minimize the effect of the index-$\sigma$ relations and
determine the radial index gradients as a function of $\sigma$. We plot index values as a
function of $r/r_e$ in 5 bins, for 8 separate bins of $\sigma$. The bins are chosen to
maintain a large number of objects in each bin and gradients are then derived by
weighting each point by $1/\sqrt n$ where n is the number in the bin. The values of the
radial index gradients and their variation with $\sigma$ are shown in 
Table~\ref{tab:radgrad}.

We use the radial index gradients to correct our index values for aperture effects.
Firstly, we use the measured value of $\sigma$ for a given galaxy to determine the value
of the radial index gradient $\Delta I$:

\begin{equation}
\Delta I = (\sigma_{200}-1) \; \frac{d \Delta I}{d \sigma_{200}} + \Delta I_{200}
\end{equation}    

\noindent where $\sigma_{200}$ is the velocity dispersion in units of $200\;\rm km\,s^{-1}$, and $\Delta I_{200}$ is the value of the index gradient for $\sigma = 200\;\rm km\,s^{-1}$. Values for $d \Delta I / d \sigma_{200}$ and $\Delta I_{200}$ are given in Table~\ref{tab:radgrad}.

This value of the radial index gradient is then used with the measured value of $r_e$
of the galaxy to correct the index value to the equivalent radius, $r = \rm r_e/2$ or
$\rm r_e/4$:

\begin{equation}
I_c = I + \Delta I \; \log(r\,r_e/1\farcs5)
\end{equation}    

\noindent where $I_c$ is the corrected index, $I$ is the measured index value, $r_e$ is the effective radius of the galaxy in arcsec and $r$ is the standard radius expressed as a fraction of $r_e$ (0.5 or 0.25 here).

The use the the effective radii provided in SDSS (the Petrosian half-light radius) to
effect the aperture correction has one caveat. That is, that these radii are not seeing
corrected. Therefore for galaxies with a small angular diameter $r_e$ is over-estimated.
Median r-band seeing for SDSS imaging is $1\farcs4$ (Adelman-McCarthy et al. 2007, fig.~4)
and the median effective diameter for our sample is $5^{\prime\prime}$. We believe, however, that this is not
a serious problem because the seeing also alters the light entering the spectroscopic
fibre. The considerable agreement we find in values for the indices as a function of
$\sigma$ with other studies (see below) reinforces this view.

The second difference from Paper~1 is the removal of one index, G4300, from the fitting procedure used to derive the
age, metallicity and $\alpha$--enhancement as a function of $\sigma$ and environment.
This was done because the index seems less well modeled than previously thought.

\section{Results}

Here we will use the index values to derive various evolutionary parameters as a function
of $\sigma$, environment and galaxy radius. Before that, however, we briefly evaluate
various trends seen in the fully $\sigma$ and aperture corrected index values.

\subsection{Index values as a function of velocity dispersion}

In Table~\ref{tab:grad} we show the corrected index values as a function of $\sigma$.
Because the values refer to indices aperture corrected to $r_e/2$, they are not
directly comparable to those of Paper~I, where corrections were made to $r_e/10$.
Nonetheless, most indices show similar behaviour to those of Paper~I. We briefly note
here some of the larger differences seen in important indices (we refer to gradients
expressed as $dI/d\log(\sigma)$ as $a_1$ and those as $d\log I/d\log(\sigma)$ as $A_1$).

{\bf C4668}: The gradient of $a_1 = +4.4$ much larger than that of Paper~I ($+1.8$). This
value is closer to the value of $+5.2$ found by  Nelan et al. (2005, hereafter N05). {\bf
H$\beta$}: The gradient of $-1.0$ ($A_1=-0.24$) is shallower ($-1.7$ in Paper~I). This value is
in excellent agreement with Bernardi et al. (2003, hereafter B03) ($A_1=-0.24$) and N05 (
$a_1 = -1.2$). {\bf Fe5015}: The gradient of $a_1=-1.5$ in Paper~I contrasts to the
present value of $+1.6$. This is much more consistent with N05 who find $a_1=+1.0$. {\bf
Mg2}: The value of $a_1=0.22$ is similar to that of Paper~I, but is now more consistent
with both B03 and Kuntschner et al. (2002). {\bf Mgb}: The gradient of $a_1 = 3.3$ (
$A_1=0.37$) compares with 3.7 in Paper~I. This is closer to that of B03 ($A_1=0.32$) and
N05 ($a_1 = 3.2$). {\bf Fe5270}: In Paper~I a null gradient was found. The present value
of $a_1=+0.62$ is consistent with N05 who find an identical value.

We note, that globally, the new index gradients are much closer to those derived by N05
despite the fact that these authors aperture corrected index values to a fixed physical
radius, rather than to a fixed fraction of $r_e$ as done here.

\subsection{Index values as a function of radius}

The spatial gradients we measure here describe the \emph{mean index
value in apertures of varying radii}, at fixed $\sigma$. This is in
contrast with `true' spatial index gradients, which are measured in
increasing annuli projected on the galaxy. As a result, the values we
measure are smaller in magnitude than the true gradients.  Our values,
which we give in Table ~\ref{tab:radgrad}, are, however, directly
applicable to aperture correction.

All the narrow line indices, with the exception of HK, show a radial gradient. The
gradients are negative except for the hydrogen line indices 
and the $\rm 4000\AA$ break, B4000. Additionally, some indices show a well
defined trend of index gradient with $\sigma$ (these can be quickly identified in the
last column of Table~\ref{tab:radgrad}). In all cases (except CN1 and CN2)
the sense of this variation is that the index gradient becomes less steep with increasing
$\sigma$. In some cases, including $\rm H\beta$, a significant index gradient at low
values of $\sigma$ disappears completely for $\sigma > 250\;\rm km\,s^{-1}$. We find no
dependence on the radial index gradients with density of environment at fixed $\sigma$.
The fact that some indices show gradients which decrease with increasing $\sigma$ does 
not necessarily imply that some process has acted to mix the stellar populations 
in more massive systems. We return to this below.

\begin{table}
\centering
\begin{tabular}{l r r c}
\hline\hline
Index &  $\Delta I_{200}\quad\quad$ & $d \Delta I / d \sigma_{200}$ & S/N\\
\hline
     CN1[mag] &  $ -0.019 \pm 0.002$ & $ -0.005 \pm 0.008$ & $ 0.6$\\
     CN2[mag] &  $ -0.018 \pm 0.002$ & $ -0.010 \pm 0.009$ & $ 1.1$\\
       Ca4227 &  $ -0.113 \pm 0.014$ & $  0.060 \pm 0.065$ & $ 0.9$\\
        G4300 &  $ -0.012 \pm 0.031$ & $  0.379 \pm 0.149$ & $ 2.5$\\
       Fe4383 &  $ -0.423 \pm 0.040$ & $  0.600 \pm 0.188$ & $ 3.2$\\
       Ca4455 &  $ -0.121 \pm 0.017$ & $ -0.008 \pm 0.081$ & $ 0.1$\\
       Fe4531 &  $ -0.188 \pm 0.030$ & $  0.203 \pm 0.142$ & $ 1.4$\\
        C4668 &  $ -1.039 \pm 0.057$ & $ -0.561 \pm 0.282$ & $ 2.0$\\
     H$\beta$ &  $  0.085 \pm 0.021$ & $ -0.488 \pm 0.105$ & $ 4.7$\\
       Fe5015 &  $ -0.415 \pm 0.043$ & $ -0.401 \pm 0.201$ & $ 2.0$\\
     Mg1[mag] &  $ -0.024 \pm 0.001$ & $  0.013 \pm 0.005$ & $ 2.5$\\
     Mg2[mag] &  $ -0.028 \pm 0.001$ & $  0.013 \pm 0.007$ & $ 2.0$\\
          MgB &  $ -0.311 \pm 0.028$ & $  0.286 \pm 0.128$ & $ 2.2$\\
       Fe5270 &  $ -0.173 \pm 0.023$ & $  0.066 \pm 0.112$ & $ 0.6$\\
       Fe5335 &  $ -0.241 \pm 0.026$ & $ -0.134 \pm 0.127$ & $ 1.1$\\
       Fe5406 &  $ -0.150 \pm 0.022$ & $  0.010 \pm 0.108$ & $ 0.1$\\
       Fe5709 &  $ -0.053 \pm 0.015$ & $ -0.052 \pm 0.074$ & $ 0.7$\\
       Fe5782 &  $ -0.103 \pm 0.015$ & $  0.058 \pm 0.071$ & $ 0.8$\\
          NaD &  $ -0.627 \pm 0.037$ & $ -0.546 \pm 0.173$ & $ 3.1$\\
    TiO1[mag] &  $ -0.000 \pm 0.001$ & $  0.004 \pm 0.002$ & $ 1.9$\\
    TiO2[mag] &  $ -0.008 \pm 0.001$ & $  0.002 \pm 0.002$ & $ 0.8$\\
        B4000 &  $  0.017 \pm 0.002$ & $ -0.013 \pm 0.007$ & $ 1.9$\\
           HK &  $  0.001 \pm 0.002$ & $ -0.028 \pm 0.012$ & $ 2.4$\\
   H$\delta$F &  $  0.265 \pm 0.030$ & $ -0.265 \pm 0.144$ & $ 1.8$\\
   H$\gamma$F &  $  0.290 \pm 0.031$ & $ -0.558 \pm 0.154$ & $ 3.6$\\
\hline
\end{tabular}
\caption{Radial index gradients as a function of velocity dispersion, $\sigma$. 
The radial index gradients are expressed as, $\Delta I = \frac{dI}{d \log(r/r_e)}$,
where $I$ is the value of the index.
The third column gives values for the dependence of the radial index gradients 
on $\sigma$ in units of $200\;\rm km\,s^{-1}$, $\sigma_{200}$. 
$\Delta I_{200}$ is the value of the index gradient at $\sigma=200\;\rm km\,s^{-1}$.
These parameters have been used in the aperture correction of all indices.
Most indices show a well-defined radial gradient, but rather few show a convincing trend of this gradient on $\sigma$. The last column shows the ratio of the gradient in the third column and its error and so is an estimate of the statistical significance of the gradient as a function of $\sigma$. For those indices whose radial gradients show little trend with $\sigma$, the value of $\Delta I_{200}$ is a good measure of the radial gradient for galaxies of any $\sigma$.}
\label{tab:radgrad}
\end{table}

\begin{table}
\centering
\begin{tabular}{l r r}
\hline\hline
Index & $a_1=dI/d\log(\sigma)$ & $a_0\quad\quad\;$\\
\hline
    CN1 [mag] & $  0.209 \pm  0.002$ & $ -0.413 \pm  0.005$\\
    CN2 [mag] & $  0.224 \pm  0.003$ & $ -0.416 \pm  0.006$\\
       Ca4227 & $  0.400 \pm  0.020$ & $  0.190 \pm  0.046$\\
        G4300 & $  1.118 \pm  0.047$ & $  2.758 \pm  0.106$\\
       Fe4383 & $  1.759 \pm  0.061$ & $  0.699 \pm  0.136$\\
       Ca4455 & $  0.628 \pm  0.024$ & $ -0.172 \pm  0.053$\\
       Fe4531 & $  1.184 \pm  0.044$ & $  0.544 \pm  0.100$\\
        C4668 & $  4.392 \pm  0.083$ & $ -3.516 \pm  0.187$\\
     H$\beta$ & $ -1.033 \pm  0.032$ & $  4.032 \pm  0.073$\\
       Fe5015 & $  1.604 \pm  0.061$ & $  1.242 \pm  0.137$\\
    Mg1 [mag] & $  0.145 \pm  0.002$ & $ -0.217 \pm  0.004$\\
    Mg2 [mag] & $  0.216 \pm  0.002$ & $ -0.236 \pm  0.005$\\
          Mgb & $  3.305 \pm  0.057$ & $ -3.325 \pm  0.128$\\
       Fe5270 & $  0.617 \pm  0.038$ & $  1.427 \pm  0.085$\\
       Fe5335 & $  0.922 \pm  0.039$ & $  0.495 \pm  0.087$\\
       Fe5406 & $  0.563 \pm  0.032$ & $  0.384 \pm  0.072$\\
       Fe5709 & $ -0.148 \pm  0.021$ & $  1.206 \pm  0.047$\\
       Fe5782 & $  0.348 \pm  0.026$ & $ -0.015 \pm  0.059$\\
          NaD & $  4.432 \pm  0.049$ & $ -6.345 \pm  0.110$\\
   TiO1 [mag] & $  0.029 \pm  0.001$ & $ -0.032 \pm  0.002$\\
   TiO2 [mag] & $  0.045 \pm  0.001$ & $ -0.026 \pm  0.002$\\
        B4000 & $ -0.128 \pm  0.003$ & $  0.848 \pm  0.006$\\
           HK & $ -0.064 \pm  0.003$ & $  1.003 \pm  0.008$\\
   H$\delta$F & $ -1.600 \pm  0.044$ & $  4.032 \pm  0.099$\\
   H$\gamma$F & $ -2.530 \pm  0.046$ & $  4.333 \pm  0.104$\\
\hline
\end{tabular}
\caption{Index values as a function of velocity dispersion, $\sigma$. 
The indices have been aperture corrected to a standard radius of $r_e/2$. Straight line fit
parameters are given as the gradient, $a_1=dI/d\log(\sigma)$, and intercept, $a_0$, of plots
of index versus $\log\sigma$.}
\label{tab:grad}
\end{table}

\subsection{Stellar population trends}

We now make use of the index values to derive the age, metallicity and $\alpha$-
enhancement of the galaxy population in our sample. We repeat the multiple linear
regression procedure described in Paper~I (to which the reader is referred for a detailed
description). We briefly summarize the procedure here.

Firstly, because our index values are not calibrated to the Lick system (due to the lack
of Lick standard star spectra in SDSS) we consider index variations relative to the mean
value at a $\sigma$ of $\sigma=200\;\rm km\,s^{-1}$. By working with these differential
index values we avoid both the problem of absolute calibration to the Lick system and
potential problems in the absolute calibration of the SSP models. We therefore perform a
multiple linear regression according to equation 5 of Paper~I.
The linear regression is performed on the whole sample and on 2 subsets of environmental
density, $1/r_{5} \leq 0.5$ (typical of the 'field') and $1/r_{5} \geq 1.5$ (more typical
of a cluster). We also perform the analysis on indices aperture corrected to 2 different
radii, $r_e/2$ and $r_e/4$, to investigate radial trends within the individual galaxies.

\begin{figure}
\includegraphics[angle=0,width=0.45\textwidth]{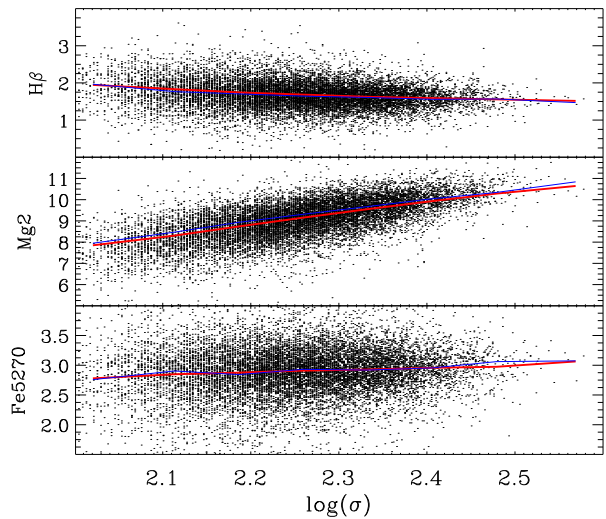}
\caption{A posteriori comparison of the models with selected indices 
(only three are shown but all indices are similarly reproduced). The 
thick red lines are the simultaneous solution to the whole set of indices 
(not fits for each index), while the blue lines trace the median 
data value.}
\label{fig:model_fits}
\end{figure}

The results of this regression analysis showed that the carbon abundance did not depend
on $\sigma$, having a constant offset as a function of environment, in contrast to Paper~
I. This difference is probably due to the better aperture correction used here and/or the
exclusion of the G4300 index. We therefore remove the explicit carbon abundance from the
regression analysis, allowing the carbon abundance to be included in the metallicity
term. The \emph{simultaneous a posteriori fit} of the model to three example indices is
compared with the median of the data in the different bins of $\sigma$, in 
fig.~\ref{fig:model_fits}.

\begin{figure}
\includegraphics[angle=-90,width=0.47\textwidth]{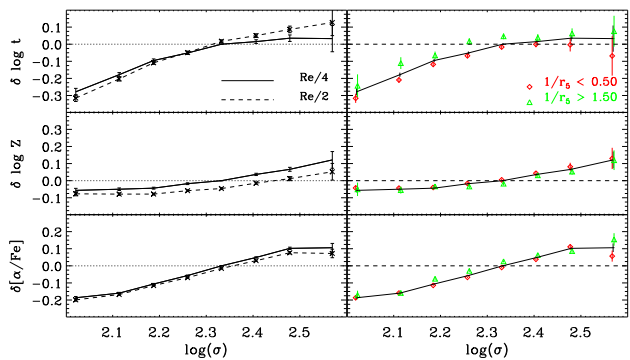} 
\caption{Age, metallicity and $\alpha$-enhancement variations as a function 
of $\sigma$. A linear regression analysis has been carried out simultaneously 
on the H$\beta$, H$\delta$, Mg1, Mg2, MgB, Fe4383, Fe4531, Fe5270, Fe5335 and C4668 indices. 
{\bf Left:} Radial variation. The two lines in each plot refer to index values corrected
to $r_e/4$ (solid line) and $r_e/2$ (dashed line).
{\bf Right:} Effect of environment. The solid line represents
the entire sample, diamonds only those objects in low density environments
($1/r_5 < 0.5$) and triangles only those in high density environments
($1/r_5 > 1.5$). Values are differences with respect to those of the entire
sample at $\sigma = 200\;\rm km\,s^{-1}$. The central $\sigma$ bins 
typically contain $> 10^3$ galaxies, whereas the highest bin contains only 16.}
\label{fig:results}
\end{figure}

In the left panel of Figure~\ref{fig:results} we show the results for the entire sample
for two different radii. For $r_e/4$ the trend of age with $\sigma$ is very similar to
that seen in Paper~I with a steady rise in age from the lowest mass systems and an
approximately constant age for galaxies with $\sigma > 230\;\rm km\,s^{-1}$. For the
larger aperture, however, the trend is slightly different, with a less pronounced
flattening towards high values of $\sigma$. For $\sigma > 300\;\rm km\,s^{-1}$ the mean
age is $\simeq 0.05$~dex older for $r_e/2$ compared to $r_e/4$. This corresponds to an
age difference of $\sim 1\;\rm Gyr$ for a galaxy of age $10\;\rm Gyr$.
The cross-over point of the 2 lines in the top
panel of fig.~\ref{fig:results} shows that galaxies with $\sigma > 180\;\rm km\,s^{-1}$
have positive radial age gradients. Neither S{\'a}nchez-Bl{\'a}zquez et al. (2007) nor
Mehlert et al. (2003) find evidence of radial age gradients in ETGs. We derive an age
gradient in massive galaxies \emph{despite the absence of a gradient of the H$\beta$
index}.

The increase of metallicity with $\sigma$ is less strong than that found in Paper~I. The gradient for the indices, corrected to $r_e/4$, for $\sigma > 160\;\rm km\,s^{-1}$ is $d\,\log(Z) / d\,\log(\sigma)\simeq 0.42$. This value is similar to that found by Nelan et al. (2005), Thomas et al. (2005) and Smith et al. (2007) but smaller than that of Kuntschner et al. (2001) and Graves et al. (2007). For $\sigma < 160\;\rm km\,s^{-1}$, however, there is no significant trend of index value with $\sigma$.  For the $r_e/2$ aperture, the metallicity is $\simeq 0.05$~dex lower so that early-type galaxies are less metal rich at larger radii. Negative metallicity gradients have also been reported by Proctor et al. (2008), Annibali et al. (2007), S{\'a}nchez-Bl{\'a}zquez et al. (2007) and Harris \& Harris (2002). There is also evidence that the metallicity gradients are steeper for more massive galaxies as found by Forbes et al. (2005). The metallicity gradient likely compensates the age gradient to remove radial gradients in indices like H$\beta$.

The trend of $\alpha$-enhancement with $\sigma$ is also slightly less steep than found in
Paper~I, with $d\,[\alpha/{\rm Fe}] / d\,\log(\sigma)\simeq 0.55$, similar to that found
by Annibali et al. (2007) but steeper than the $\sim 0.3$ found by several authors (
Thomas et al., 2005, Kuntschner et al., 2001, N05, Bernardi et al., 2006, Smith et al.,
2007). Although most similar studies refer to a smaller radius, we see that the larger
aperture, $r_e/2$, has a marginally shallower gradient. The $\alpha$-enhancement within
this larger aperture is slightly lower, with the difference being largest ($0.03$~dex)
for the most massive galaxies. This does not support the `outside-in' ETG formation scenario
 (Pipino, Matteucci, \& Chiappini, 2006). A negative $\alpha$-enhancement gradient is also seen in
the halo stars of the Galaxy (Fulbright, 2000).

In the right panel of Figure~\ref{fig:results} we show the variation of evolutionary parameters as a function of the density of environment for index values aperture corrected to $r_e/4$. In both environments the increase in age with $\sigma$ is similar, with a flattening above $\sigma \sim 200\;\rm km\,s^{-1}$. objects in dense environments ($1/r_{5} \geq 1.5$), however, are $0.087$~dex older than those in less dense environments ($1/r_{5} \leq 0.5$). This is a difference of 2~Gyr if the ages are close to 10~Gyr and is consistent with several earlier studies (Terlevich \& Forbes 2002, Kuntschner et al. 2002, de la Rosa et al. 2007, S{\'a}nchez-Bl{\'a}zquez et al. 2006). The flattening of the age-$\sigma$ relation is also slightly more pronounced for the lower density environment. Similar age trends were seen in Paper~I.

There is marginal evidence that the metallicity is lower in high density environments.
Formally the difference is $0.020 \pm 0.013$~dex. Thomas et al. (2005) also found a small
environmental dependence on the metallicity in the same sense and de la Rosa (2007) finds
a difference of $0.11$~dex between Hickson Compact Groups and the field. Other authors
have found both larger differences in the same sense (Proctor et al. 2004, Kuntschner et
al. 2002), no effect (Bernardi et al., 2006, Annibali et al. 2007) and the opposite effect
(Gallazzi et al. 2006, Mateus et al. 2007). In Paper~I no difference was found.

No environmental effect is found for the $\alpha$-enhancement, in agreement with
Kuntschner et al. (2002), Thomas et al. (2005), Annibali et al. (2007) and Gallazzi et
al. (2006). However, Proctor et al. (2004), Bernardi et al. (2006) and Lucey et al.
(2007) all find increased $\alpha$-enhancement in denser environments.

\section{Discussion and conclusions}

We find positive correlations between age, metallicity and $\alpha$-enhancement and the
velocity dispersion, $\sigma$, in ETGs. Galaxies in dense environments are $\sim 20\%$ older than those in low density environments for all $\sigma$ ($\sim 2\;\rm Gyr$ for an age of $10\;\rm Gyr$).
The trend with age flattens above $\sigma \sim 200\;\rm km\,s^{-1}$, especially for galaxies
in low density environments. We find a marginally significant trend towards higher
metallicities in low density environments but the environment has no effect on the
$\alpha$-enhancement.

Apart from the marginal metallicity difference between field and cluster the results are
very similar to those of Paper~I. There we concluded that an anti-hierarchical scenario,
in which star formation lasts longer but with lower efficiency in lower mass objects (see
Granato et al. 2004) was consistent with the data. Here the additional determination of
SSP parameters as a function of galactic radius places additional constraints on the
evolutionary scenario.

Massive ETGs ($\sigma > 180\;\rm km\,s^{-1}$) have positive radial age gradients,
negative metallicity gradients and marginally significant negative $\alpha$-enhancement
gradients. When a massive halo becomes non-linear it accretes smaller halos which started
to collapse at earlier times. 
The radial trends suggest that these halos do not contain
only gas, but also pristine stars. 
The gaseous component falls dissipatively into the potential
well of the massive (proto-)spheroid, fueling rapid star formation. 
The increase in mass
increases the rate and efficiency of star formation, driving the main correlations with
galaxy mass. The pristine stellar component of each sub-halo, however, 
being dissipationless, is deposited at a radius consistent with the angular momentum 
of the encounter. These stars, which are slightly older, more metal poor and have moderate 
$\alpha$-enhancement will therefore be spread over larger radii. 
At early times, rapid gas-rich
mergers lead to an almost monolithic formation, at later times mergers become
increasingly ``dry''. Very similar scenarios have been proposed to explain both the bi-
modal metallicity distribution of globular clusters and the greater radial scale length 
of metal-poor relative to metal-rich globular clusters in elliptical galaxies 
(C\^ot\'e, Marzke \& West, 1998, Bekki et al. 2008). 
Our results imply that the metal-poor
globular cluster population in ETGs should be older, less metal-rich and slightly less
$\alpha$-enhanced than the metal-rich clusters.

Because the age difference at larger radii is actually the luminosity weighted SSP
equivalent age in a larger aperture (not an annulus), the value of 0.05 dex, $\sim 1\;\rm Gyr$, is a lower limit to the real age difference at larger radii. This time difference limits the assembly redshift of massive ETGs simply due to the lack of time to accommodate the formation of stars in the lower mass halos. Our limit
translates into an upper limit to the assembly redshift of massive ETGs of
$z \lesssim 3.5$; in which case the stars in low mass halos formed at $z\sim 10$, for a
standard cosmology ($H_0=70$, $\Omega_m = 0.3$, $\Omega_{\Lambda}=0.7$). This also provides an estimate of the star formation rate in the
assembled spheroid. If the final stellar mass were $10^{12}\;\rm M_{\odot}$ then stars
must have formed at a rate
$\sim 10^{12}\;\rm M_{\odot}/1\;\rm Gyr \sim 10^3\;\rm M_{\odot}\,yr^{-1}$.

We conclude by stressing the statistical nature of our results. Because a galaxy's 
velocity dispersion is a function of both the halo mass and virialization redshift, 
variations in these parameters may render small samples insensitive to the trends we find. 

The catalogue on which this article is based can be found at, {\tt www.mrao.cam.ac.uk/$\sim$bn204/galevol/clemensetal08.html}.

\section*{Acknowledgments}

Funding for the SDSS and SDSS-II has been provided by the Alfred P. Sloan Foundation, the Participating Institutions, the National Science Foundation, the U.S. Department of Energy, the National Aeronautics and Space Administration, the Japanese Monbukagakusho, the Max Planck Society, and the Higher Education Funding Council for England. The SDSS Web Site is http://www.sdss.org/.
We acknowledge a financial contribution from contract ASI-INAF I/016/07/0.

\bsp

\label{lastpage}

\end{document}